# REVERSE ENGINEERING ONTOLOGY TO CONCEPTUAL DATA MODELS

Haya El-Ghalayini, Mohammed Odeh, Richard McClatchey & Tony Solomonides

Centre for Complex Cooperative Systems, CEMS Faculty, University of the West of England (UWE),
Coldharbour Lane, Frenchay, Bristol BS16 1QY, United Kingdom
Email: Haya2.Elghalayini@uwe.ac.uk

**ABSTRACT**. Ontologies facilitate the integration of heterogeneous data sources by resolving semantic heterogeneity between them. This research aims to study the possibility of generating a domain conceptual model from a given ontology with the vision to grow this generated conceptual data model into a global conceptual model integrating a number of existing data and information sources. Based on ontologically derived semantics of the BWW model, rules are identified that map elements of the ontology language (DAML+OIL) to domain conceptual model elements. This mapping is demonstrated using TAMBIS ontology. A significant corollary of this study is that it is possible to generate a domain conceptual model from a given ontology subject to validation that needs to be performed by the domain specialist before evolving this model into a global conceptual model.

## 1. INTRODUCTION

Semantic heterogeneity represents a major challenge facing the integration and inter-operation of large scale data and information sources. As Sowa [1] asserts, semantics "determine how the constants and the variables are associated with things in the application domain". Also, Sheth and Larson [2] refer to the occurrence of semantic heterogeneity within databases as the "disagreement about the meaning, interpretation or intended use of the same or related data [in different databases]".

In computer-based information systems, the meaning of information is usually captured in terms of conceptual models that offer semantic terms for modelling applications and structuring information [3]. However, defining terms and mechanisms for information modelling using conceptual models requires assumptions about the problem domain and the application being modelled. Therefore, entities and relationships are established based on their semantics *in relation to the problem and application domain*. In this case, ontology identifies, within a certain domain, the concepts for modelling a world for which one would like to do computations and to explicitly specify the formal definition of semantics of the terms [4]. This research is an attempt to study the possibility of generating a domain conceptual model from a given domain ontology (i.e. reverse engineering a given ontology) by using a set of transformation rules implemented within a transformation engine component. In this paper, sections 2 and 3 introduce some background in ontologies and conceptual models. Then, section 4 presents first this transformation engine by briefly presenting conceptual modelling in general and ontology language constructs based on the BWW-model. Then, the mapping between an ontology in DAML+OIL and conceptual modelling elements is explained. Section 5 introduces the case study, namely the Ontology of the TAMBIS[1] project. Finally, discussion and conclusion are presented in sections 6 and 7, respectively.

## 2. ONTOLOGIES

Since the early 1990s, ontologies have been used within several artificial intelligence research projects to facilitate the sharing and reuse of knowledge. Lately, ontologies have become the focus for research in several other areas, including knowledge engineering and management, information retrieval and integration, digital libraries, the semantic web, and e-commerce. A popular definition of ontology was proposed by Gruber [5] as "a formal, explicit specification of a shared conceptualization". The term *formal* refers to the fact that ontology should be machine readable; *explicit* means that the type of concepts used, and the constraints on their use are explicitly defined; *shared* reflects that ontology should capture consensual knowledge accepted by the communities; and *conceptualization* refers to an abstract model of phenomena in the world by having identified the relevant concepts of those phenomena. In conclusion, this may be summarized as expressing knowledge in machine-readable form in order to permit knowledge exchange between heterogeneous environments. Guarino [6] classifies ontologies into four levels: top-level, domain, task, and application level ontologies. Furthermore, Jurisica, et al. [4] classifies the concepts used for knowledge representation into four broad ontological categories: static, dynamic, intentional and social. Developing a static ontology is the most prevalent ontology-based activity – such as taxonomies or controlled vocabularies [7, 8] – and its main aim is to standardize the domain concepts to enable information sharing and system cooperation.

Depending on the application domain, ontologies can be represented using different approaches, for example as a hierarchy of concepts for some applications while others requiring more complex constraints. In fact, the greater the requirements, the more complicated and expressive knowledge the representation language has to be. Knowledge representation (KR) is used to marry semantic information in a domain of the world with the functionality of intelligent reasoning systems. KR languages have developed from semantic networks and frame-based systems to include description logic. In our research, we are utilizing DAML+OIL which, like conceptual modelling languages, allows the specification of classes of objects (i.e. entity types) and properties (i.e. relationships). Classes in DAML+OIL can be defined either directly by their explicit names or by composition, as a class expression built using a variety of constructors. Classes can be combined using conjunction, disjunction, and negation constructors. They can be either *primitive* or *defined classes*. Primitive classes are explicit classifications, whereas the defined classes are expressions capturing the necessary and sufficient conditions for membership of a class. Properties are explicit descriptions of the attributes that determine class membership. A set of axioms can be used to assert different relationships such as *subsumption*, *hasClass/toClass,* or *disjointness of classes* [9].

## 3. CONCEPTUAL MODELLING

The task of Conceptual Modelling (CM) plays a crucial role in the process of information systems development. Conceptual models translate and specify the main data requirements of the user requirements in an abstract representation of selected semantics about some aspects of a real-world domain. Systems analysts seek to capture and represent all relevant problem domain entities and their relationships. In addition, conceptual modelling languages and notations were introduced to represent conceptual models using a collection of modelling elements. In general, any conceptual model can be considered as a tuple: CM= (E, R, A, C), where E stands for Entities, R for Relationships, A for Attributes, and C for Constraints including semantic integrity constraints.

## 4. REVERSE ENGINEERING ONTOLOGY TO CONCEPTUAL MODEL

The proposed architecture of our approach is depicted in figure 1. The main component in this architecture is the Transformation Engine which consists of two sets of Transformation Rules. The first set of rules is responsible for reverse engineering domain ontology to corresponding domain conceptual data model. This generated model may act as a domain meta-model that describes the basic model elements and the relationships between them as well as their semantics in a certain domain.

The second set of transformation rules is for constructing the global conceptual schema to be derived as an instance of the domain conceptual model and modified based on the anticipated local conceptual schemas of the underlying data sources. The contribution of our approach is two-fold. Firstly, it automates (to a certain extent in addition to domain specialist validation) the construction of a domain conceptual model from domain ontology. Secondly, since the domain ontology unifies the concepts and their relations, it facilitates the integration of heterogeneous, autonomous, and distributed data sources in reconciling the semantic heterogeneity, which is considered a major obstacle in data integration approaches. The second layer ($2^{nd}$ rule set)-constructing the global conceptual schema from the conceptual model-is beyond the scope of this paper.

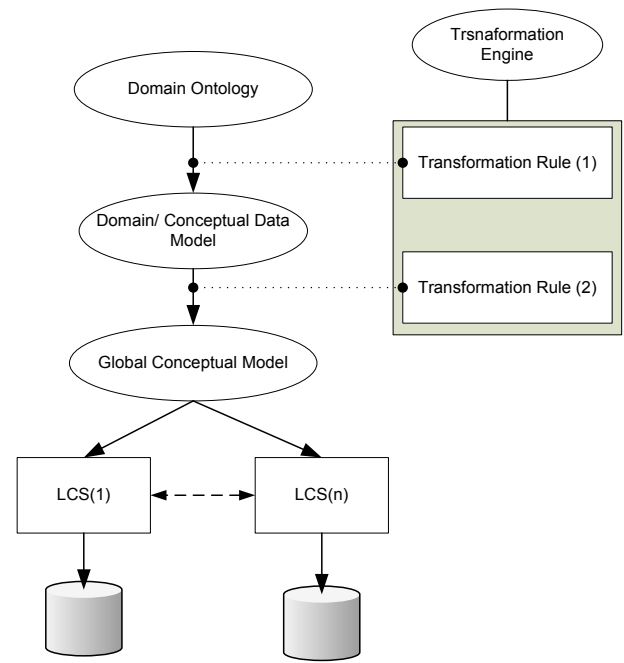

**Figure 1**: General architecture of the proposed approach

## 4.1 THE TRANSFORMATION ENGINE

In our approach, the first set of Transformation Rules [TR(1)] utilizes the ontological model provided by Wand and Weber, which is known as Bunge-Wand-Weber model (BWW). Wand and Weber studied the ontology-branch of philosophy-as a foundation for understanding the meaning of constructs in modelling languages in order to achieve an adequate conceptualization of a certain domain [10]. They have extended an ontology presented by Bunge and applied it to modelling grammars. Their fundamental hypothesis is that any modelling grammar must be able to represent all things in the real world that might be of interest to users of information systems. Thus, they defined a set of abstract concepts (constructs) needed for the theoretical foundation while in the process of information systems analysis and design. The complete representation of these constructs describes the structure and behaviour of a system.

Briefly, we now explain the fundamental structural concepts in BWW used for modelling information systems. The world consists of *substantial things* (individuals) that possess *properties*. A property can be either *intrinsic* - a property of an individual thing, or *mutual* - a property that is meaningful for two or more things. Things can combine to form a composite thing where its properties can be *emergent* and not possessed by any component. A *class* is the set of things that have a common property, a *kind* is the set of things that have two or more common properties and *a natural kind* is a kind with restricted laws over properties. A set of *attributes* used to describe a set of things with common properties is called a *functional schema*. Laws exist to restrict the possible *states* of a thing. The behavioural concepts describe how things have changed or may change. The *event* is a change of state. The *interaction* among things is captured by the concept *coupling* or *bounding* – which is represented by a mutual property that indicates at least one of the things acting another other [11, 12].

This research focuses on interpreting the ontology and conceptual language elements using the basic structural constructs defined in BWW model. Note that the TR(1) builds on Baclawski's work (which is the inverse of our approach) that utilizes UML class diagrams to build complex DAML ontologies [13]. As a result, applying transformation mapping on any domain's ontology generates a conceptual data model (which may be considered as a domain meta-model) that describes the domain entities, relationships between these entities with all [2] domain restrictions being satisfied as shown in figure 2.

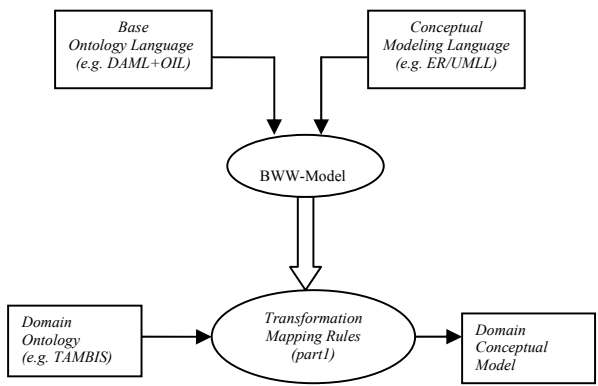

**Figure 2**: Proposed framework for generating domain conceptual model from domain ontology

## 4.2 MAPPING ONTOLOGY CONSTRUCTS TO CONCEPTUAL MODELLING CONSTRUCTS

The mapping of the modelling language constructs into ontological concepts is called the interpretation mapping [10]. Analysis of this mapping identifies the language elements that either have no ontological constructs *(construct excess)* which leads to an ontologically meaningless model or have multiple ontological constructs *(construct overload)* which leads to ambiguous models and results in a defective information system [11]. Different researchers have applied the BWW-model on different conceptual modelling grammars [14, 15].

In addition, ontology languages focus on explicit and formal representation of concepts and their relationships in a hierarchical structure within the domain of interest. McGuinness describes three main *types of properties* needed for ontology [16]. Accordingly, an ontology can be defined as ON tuple: (C, P, R, I, A) where C stands for Concepts, P for Properties, R for Restrictions, I for Individuals, and A for Axioms. According to BWW-model we map the concept in the domain ontology to a class, a kind or a natural kind: in most cases it will be mapped to a natural kind since the concept specifies the semantics of a thing and its relations with other concepts. *Properties* in domain ontology are mapped to properties in the BWW-model. Restrictions and axioms are mapped to the laws in BWW-model. Finally, an individual corresponds to a thing. We noticed that further work is needed to define a set of rules and guidelines about how to specify concepts in the domain ontology. This can help in developing and specifying reusable domain ontologies as a basis for developing different application within the same domain. This, however, may assist in resolving the semantic heterogeneity between their related data sources. In this paper, we propose an initial set of mapping rules to generate a conceptual model from the domain ontology by using the BWW ontological constructs to match between elements. However, concepts used in knowledge representation languages in a machine readable form, i.e. ontology, are very close to those used to represent data in conceptual models; both models have much in common. Although the aim of each model is different, reverse engineering the domain ontology assists in developing the domain conceptual model, where the latter is a high-level model which contains high-level domain entities and their basic relationships with other major domain entities. The rules below briefly summarise the transformation rules used in this research and are part of TR(1) component in figure 1.

Rule 1: *BWW-thing* represents both an individual in Ontology Language (OL) and an *entity* in the conceptual model (CM). So, we map *individual* to an *entity* or object in CM.

Rule 2: Since the *class* in OL and entity type in CM are both represented by *BWW-class, kind or natural kind*, then we can say that every *named class* presented in OL is mapped to an *entity type* under *certain constraints*.

Rule 3: *BWW-intrinsic property* represents the *daml:DatatypeProperty* in OL and *entity's attribute* in CM. Then, every *daml:DatatypeProperty* used in defining a class is assigned to an *entity-attribute* of the studied class in addition to applying the *global/local property constraints* associated with this property .

Rule 4: BWW-mutual property represents both the *property* in OL and the *relationship* in CM. The super/subclass relation in OL can be defined as *a BWW-natural/sub-kind relation* that will be mapped to *a*

*generalization/specialization relation between super/sub-entity types*. As every class in OL is mapped to an entity type in CM, then every property used in the class being studied is mapped to a *relationship* between two entities by satisfying the needed constraints. In our mapping rules, there is no one-to-one mapping between a *property* in OL and a *relationship* in CM because properties are considered first-class elements in OL (i.e. *daml:objectProperty* can exist without specifying any classes that it might relate to) [13], but this is not the case in conceptual models. Therefore, we believe that it is sufficient to map only properties that are related to an identified and extracted class (satisfying both local and global constraints) in the ontology being studied.

Rule 5: *BWW-law property* represents both the *property constraints (i.e. axioms or restrictions)* in OL and the *semantic constraints* in CM. Thus, *every property constraint* used in OL class will be translated to *a relationship constraint* that restricts the kind of the relation, number of entities or entity type of this relationship. For example, *daml:intersectionOf, daml:unionOf* are mapped to *{and}* and *{or}* constraints. Also, *daml:hasClass (existential quantifier)* and *daml:toClass (universal quantifier)* determine the target class of the relationship with minimum cardinality **0** and maximum cardinality **n**. However, we add an *{exclusive}* constraint on the relationship to distinguish between *daml:toClass and daml:hasClass*. For instance, if we say that class of *proteins* can *bind* to *some DNA*, i.e. *protein* can also bind to other things {i.e. *Protein: hasClass binds DNA*}. Alternatively, the class of proteins which only bind to DNA and not other things {i.e. *Protein: toClass binds DNA*} [17]. Moreover, *cardinality* restricts the number of objects of the target entity type that can participate in a role of a relationship.

Rule 6: Some OLs have a direct representation for the whole/part or composition relation but others do not have such representation such as DAML+OIL. A *composition relation* is a mutual property between a composite and its components where the existence of the component depends on the composite. Thus, the proposed transformation rule here is map such property in ontology as a composition relation in CM if a component-class which is connected to a whole-class via a relationship must not be connected to any other class. However, this does not rule out the case that is may be a case of an aggregation rather than a composition.

## 5. CASE STUDY: THE TAMBIS ONTOLOGY

TAMBIS is considered the first ontology-based information integration system to be used in bioinformatics. The aim of the project is to allow researchers to retrieve biological information from different data sources by forming queries based on the ontology. Then, TAMBIS maps these queries to requests to the appropriate bioinformatics data sources. As a result, the biologist can easily form biological queries within a friendly graphical user interface.

The TAMBIS ontology was originally described in GRAIL and it has been migrated to OIL and then to DAML[3]. One of the main problems facing genomics data integration is related to multiple representations of the data semantics within a set of sources. For example, the same *gene* may be represented as a *genetic map locus* in one data source, an aggregation of multiple individual *exon* entries in another data source or a *set of EST sequences* in yet another one. In this case, ontologies become an integral part of bioinformatics since they encourage a common vocabulary for describing complex and evolving biological knowledge [17, 18] and can be used as a common access to diverse information repositories. The transformation engine-the first set of rules that is responsible for reverse engineering domain ontology to corresponding domain conceptual data model-has been implemented according to the rules mentioned above and was executed using the TAMBIS ontology. A snapshot[3] of the generated conceptual data (class) model is presented in figure 3 using UML. Although the results of applying the transformation rules in section four have reflected upon the main entities in the problem domain, their attributes, relationships, and some constraints, this work will need to be validated by domain specialist to assess the correctness and quality of the identified attributes, relationships, and constraints.

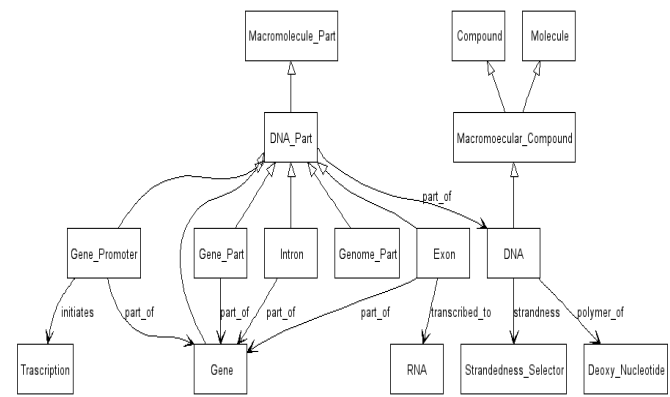

**Figure 3** – Partial output from the transformation rules

## 6. DISCUSSION

Similarities do exist between conceptual data models and ontological models with respect to abstracting and modelling the domain of discourse. But, their purposes are different. For example, class (concept) with entity-type; individuals with entities; "is-a" relation with generalization/specialization. In general, the incompatibilities between them can be summarized as follows:

Motivation: one of the main goals of ontology is to standardize the semantics of existing concepts within a certain domain. While an ontology represents knowledge that formally specifies a shared/agreed understanding of the domain of interest, conceptual models describe the structural requirements for the later storage, retrieval, organization, and processing of data in information systems such that the integrity of the data is preserved.

Usage: ontology plays a significant role at run-time to browse ontology concepts to form semantically correct queries, and perform some advanced reasoning tasks [6]. So, ontology is sharable and exchangeable at run-time, while conceptual data models are off-line model diagrams [19] and their queries are usually to retrieve a collection of instance data [20].

Evolution: generally an ontology (based on description logics) is a logical and dynamic model that can deduce new knowledge relations from the stored ones, or check for its consistency. However, conceptual models are static and are explicitly specified at design, but their semantic implications might be lost at implementation-time.

Model Elements: in ontology, elements can be expressed either by their names or as boolean expression in addition to using axioms such as cardinality/type restrictions, or domain/range constraints for classes or properties. On the other hand, conceptual model are concerned with the structure of data in terms of entities, relationships and a set of integrity constraints. For example primary key and functional dependences play very important roles within databases, but this is not always the case in the ontology since it concentrates more on how the concepts are semantically interrelated.

Meta-levels: unlike conceptual modelling languages, knowledge representation languages do not differentiate between the meta-levels. For example, KR languages provide classes that can have instances and those instances may also be classes. As a result, KR languages include modelling, meta-modelling, meta-meta-modelling etc capabilities at the same time [13].

In this paper, we have introduced a new approach to mapping a given ontology to a corresponding conceptual data model based on the ontological theory provided by BWW-model. Wand and Weber adapted Bunge's ontology to define a set of ontological constructs to present real-world phenomena in conceptual models such as thing, property, and law [21]. However, the main objective behind using BWW is assist in identifying validated CM elements based on a model that has a theoretical background such as the BWW model. And, hence this model has acted as a bridge between ontology model elements and CM elements with the significant added advantage of performing a validation step on the mapped model elements. This was in turn required to achieve adequate conceptualization, consistency, and integrity. Although the generated conceptual data model from running the TAMBIS case study has resulted in some encouraging results in terms of the discovered entities, attributes, relationships, and constraints; this model may be considered as requiring enrichment. This may be attributed to the following observations:

(1) There is no direct representation of ontology properties in the conceptual model since they are defined as first class element in the ontology language (i.e. it can exist independently of any classes). So, we are only interested in mapping the properties that are used in an ontology class with both defined domain and range.
(2) Although DAML+OIL provides *daml:hasClass* and *daml:toClass* for ontology reasoning purposes, we differentiate between them by restricting the *daml:toClass* by the proposed {excusive} constraint as raising an indicator as an alarm for the domain analyst to validate incompatible relations having this property.

(3) If we define a class A as a *subclass of* B (primitive definition) or A as *sameClassAs* B (defined definition) in the ontology, then both will be represented as a generalization/specialization relationship between B and A. In DL, equivalence class can be reduced to the subsumption relation.

(4) Aggregation is a special type of relationship that describes the entity-type as a composite of number of components. In DAML+OIL, the aggregation property cannot be expressed by the language elements explicitly. So it has to be described as any other *objectProperty* usually with the phrases such as *part-of* or *has* or *contains*. This may be still as a shortcoming in our approach as a human intervention is still required to validate the generated relation as being a correct aggregation or not.

## 7. CONCLUSION

This research aims to study the possibility of generating a domain conceptual model from a given domain ontology with the vision of growing this generated conceptual data model into a global conceptual model integrating a number of existing data and information sources. A significant corollary of this study is that it is possible to generate a domain conceptual model from a given ontology subject to some validation that needs to be performed by the domain specialist before growing this model into a global conceptual one. Furthermore, the architecture adopted in the transformation approach is both component-based and extensible. For instance, the BWW model has been used as model that has some theoretical basis to identify validated CM elements from a given ontology. This component, however, can be replaced by another component that has a different form of validation basis. In addition, this study has led us to identify some incompatibilities and shortcomings in DAML+OIL ontologies when transformed into conceptual data models as discussed in the previous section. Hence, this has paved the ground for a more robust future work to identify and specify what needs to be implemented in an ontology language to facilitate the generation of highly accurate, reliable, and less dependent on human validation conceptual data models. A further key challenge remains in the development of further approaches to grow the generated conceptual data model into a global data model to integrate heterogeneous data sources.

**Footnotes**

[1] TAMBIS (Transparent Access to Multiple Biological Information Source) system is a result of research collaboration between the departments of computer science and biological sciences at the University of Manchester in England.

[2] the ultimate objective is to satisfy all domain constraints but it might not be possible to achieve it.

[3] Experimental re-write of the TAO done in OIL in DAML+OIL file is available on
http://www.cs.man.ac.uk/~stevensr/tambis-oil.html